# Joint Planning of Distributed Generations and Energy Storage in Active Distribution Networks: A Bi-Level Programming Approach

Yang Li [a,*], Bo Feng [b], Bin Wang [a], Shuchao Sun [b]

[a] School of Electrical Engineering, Northeast Electric Power University, Jilin 132012, China

[b] State Grid Chaoyang Power Supply Company, Chaoyang 122000, China

* Corresponding author. E-mail address: liyang@neepu.edu.cn (Y. Li).

**Abstract**—In order to improve the penetration of renewable energy resources for distribution networks, a joint planning model of distributed generations (DGs) and energy storage is proposed for an active distribution network by using a bi-level programming approach in this paper. In this model, the upper-level aims to seek the optimal location and capacity of DGs and energy storage, while the lower-level optimizes the operation of energy storage devices. To solve this model, an improved binary particle swarm optimization (IBPSO) algorithm based on chaos optimization is developed, and the optimal joint planning is achieved through alternating iterations between the two levels. The simulation results on the PG & E 69-bus distribution system demonstrate that the presented approach manages to reduce the planning deviation caused by the uncertainties of DG outputs and remarkably improve the voltage profile and operational economy of distribution systems.

**Nomenclature**

| | | | |
|---|---|---|---|
| $v$ | Actual wind speed | $E_t^{WT}$ | Expected outputs of WT |
| $v_{in}$ | Cut-in wind speed | $E_t^{PV}$ | Expected outputs of PV |
| $v_{out}$ | Cut-out wind speed | $C_f$ | Present coefficients of WT |
| $v_*$ | Rated wind speed | $C_g$ | Present coefficients of PV |
| $\iota$ | Shape factor | $C_E$ | Present coefficients of storage |
| $\gamma$ | Scale factor | $C_{wd}$ | Investment costs of WT per kW |
| $P^{WT}$ | Output of WT | $C_{pv}$ | Investment costs of PV per kW |
| $P_*$ | Rated output power of WT | $S_{wd}$ | Total capacities of WT |
| $r$ | Actual light intensity | $S_{pv}$ | Total capacities of PV |
| $r_{max}$ | Maximum light intensity | $C_{ST}^{INSE}$ | Investment cost of energy storage related to rated power |
| $\lambda_1, \lambda_2$ | Shape factors | $C_{ST}^{INSS}$ | Investment cost of energy storage related to installation capacity |
| $\xi$ | Solar irradiance | $E^{ST}$ | Rated power of energy storage |
| $\eta_m$ | Maximum power point tracking | $S^{ST}$ | Installation capacity of energy storage |
| $A_{pv}$ | Radiation area of this PV | $z$ | Maintenance cost coefficient of WT and PV related to the rated power |
| $\eta_{pv}$ | Conversion efficiency | $P_{wd,sh}$ | Expected outputs of WT of hour $h$ in season $s$ |
| $\theta$ | Solar incident angle | $P_{pv,sh}$ | Expected outputs of PV of hour $h$ in season $s$ |
| $P^{PV}$ | Output of PV | $Z$ | DGs' operation cost related to the output power |
| $P_{max}^{PV}$ | Maximum output of PV | $C_{ST}^{OM}$ | Operation and maintenance cost of energy storage devices per year related to the rated power |
| $P_t^{WT}$ | WT output during period $t$ | $P$ | Active load of the system |
| $P_t^{PV}$ | PV output during period $t$ | $C_{pu,sh}$ | Time-of-use electricity price of $h$-th hour in the $s$-th season |
| $P_t^L$ | Load power during period $t$ | $P_{loss,sh}$ | Total system active power losses of $h$-th hour in the $s$-th season |
| $a(i_{a,t})$ | Probability sequence of WT output at the $t$-th sampling | $P_{e,sh}$ | Charge-discharge power of energy storage devices of $h$-th hour in the $s$-th season |
| $q$ | Discrete step size | $P_{DG,i}, Q_{DG,i}$ | Active and reactive power of DG outputs at the bus $i$ |
| $N_{a,t}$ | Sequence length of $a(i_{a,t})$ | $P_{Di}, Q_{Di}$ | Active and reactive power of load at the bus $i$ |
| $b(i_{b,t})$ | Probabilistic sequence of PV output at the $t$-th sampling | $P_{LO}, Q_{LO}$ | Total active and reactive power losses of the system without DGs |
| $N_{b,t}$ | Sequence length of $b(i_{b,t})$ | $C_{pu,h}$ | Time-of-use electricity price of $h$-th representative hour in a day |
| $U_{h,i}$ | Voltage magnitude of bus $i$ | $I_{h,ij}$ | Current of branch $ij$ |
| $U_{min}, U_{max}$ | Lower and upper bounds of the voltage of bus $i$ | $r_{ij}, x_{ij}$ and $Y_{ij}$ | Resistance, reactance and admittance of branch $ij$ |
| $I_{\max}$ | Upper current limit of any branch | $P_{h,ij}, Q_{h,ij}$ | Active/reactive power flow of branch $ij$ |
| $P_{Swing}, Q_{Swing}$ | Active and reactive power of the swing bus | | *Abbreviation* |
| $P_L, Q_L$ | Total system active and reactive power loss | DGs | Distributed generators |
| $P_{\max}^{ess}$ | Maximum charge-discharge power of energy storage | IBPSO | Improved binary particle swarm optimization algorithm based on chaos optimization |
| $S_{oc\min}, S_{oc\max}$ | Minimum and maximum allowable capacities of the energy storage | ADNs | Active distribution networks |
| $S_{oc,0}, S_{oc,24}$ | Energy storage capacities at the beginning and end of the day | WT | Wind turbine |
| $P_{loss,h}$ | Active power loss of $h$-th representative hour in a day | PV | Solar photovoltaic |
| $P_{ch,h}, P_{dc,h}$ | Charge power and discharge power of energy storage devices of $h$-th representative hour in a day | PDF | Probability density function |
| $P_{ch,\max}, P_{dc,\max}$ | Maximum charging and discharging power of energy storage devices | SOT | Sequence operation theory |
| $P_{h,i}, P_{h,j}$ | Active/reactive power injection at bus $i$ | PFV | Population fitness variance |

# 1. INTRODUCTION

Due to the increasing penetration of renewable energy resources, traditional distribution networks are evolving from passive networks to active distribution networks (ADNs). Due to the integration of distributed generations (DGs), this transition undoubtedly brings huge economic and environmental benefits, but also bring enormous challenges to the operation of the today's distribution network, such as power flow reversal [1,2]. The proper planning of DGs, such as wind turbine (WT) and solar photovoltaic (PV) generation, can reduce the line losses and the investment of the grid expansion [3]; on the contrary, an unreasonable locating and sizing strategy may lead to increased line losses, low voltage or over-voltage in some buses, and even affect the economic efficiency of distribution systems [4]. The negative impact of distributed generation sources is mainly caused by distributed generation sources' uncertainty output, which leads to the hardly reached rated power, however, the energy storage devices with decreasing cost as technology advances provide great development prospects to solve this problem [5]. Therefore, it is meaningful to consider the distributed generation sources planning of distribution networks with energy storage access.

## 1.1 literature review

The location and capacity of the distributed generation sources can be considered as a multi-objective optimization problem [6]. It is difficult to achieve the best of each target at the same time, so a trade-off between the sub-targets is necessary [7, 8]. To solve the problem of multi-objective programming, a method is presented in [9] for locating and sizing of DGs to enhance voltage stability and to reduce network losses simultaneously. First, vulnerable buses from the voltage stability point of view are determined using bifurcation analysis as the best locations to install DGs. Then, the global optimal size of DGs is determined to employ the dynamic programming search method. Similarly, the authors in [10] have proposed a multi-objective framework for simultaneous network reconfiguration and power allocation of DGs in distribution networks. The optimization problem has objective functions of minimizing power losses, operation cost, and pollutant gas emissions as well as maximizing the voltage stability index subject to different power system constraints. It is worth mentioning that the uncertainty of loads is modeled using the Triangular Fuzzy Number technique. In [11], the authors implement a multi-stage framework to handle multiple objectives in a categorical manner to simultaneously integrate DGs and energy storage devices in a distribution network.

At the same time, the location and capacity of the distributed DGs can also be considered as a single objective problem considering the actual economic benefits [12-14]. It integrates the economic indicators about DGs planning in the distribution network together to achieve the maximum benefit [15, 16]. In [17], the authors investigated microgrids generation expansion design by joint optimization method considering energy storage devices and DGs. In [18], a planning model for power distribution companies to maximize profit is proposed. The model determines optimal network location and capacity for renewable energy sources, which are categorized as independent power production and self-generation. In addition to economic indicators, other single objective functions are also covered by articles. For example, reference [19] presents a new approach for optimum simultaneous multi-DG placement and sizing based on the maximization of system load ability without violating the system constraints, and authors in [20, 21] deal with the similar problem through minimizing the power loss of the system. Unfortunately, the above references have not considered the uncertainties of DG outputs, which may result in considerable planning errors.

Due to the inherent uncertainties of DG outputs, the actual power outputs of DGs are hardly achieving (with a very low or even zero probability) their predesigned rated capacities in the planning stage by using traditional planning methodologies, so the uncertainty modeling of DGs is particularly critical for the ADN planning. The latest uncertainty handling approaches are interval-based analysis approaches, probabilistic approaches, and hybrid probabilistic approaches. In [22], under the chance-constrained programming framework, a new method is presented to handle these uncertainties in the optimal siting and sizing of DGs. First, a mathematical model of CCP is developed with the minimization of the DGs' investment cost, operating cost, maintenance cost, network loss cost, as well as the capacity adequacy cost as the objective, security limitations as constraints, and the siting and sizing of DGs as optimization variables. Then, a Monte Carlo simulation-embedded genetic-algorithm-based approach is employed to solve the developed chance-constrained programming model. In [23], the probabilistic power flow method based on the point estimate method was introduced to handle the uncertainties of DGs and load.

In addition, coupled with energy storage the DG system can perform a 'peak shaving' function and maintain the power output requirement properly, resulting in a lower core engine power rating and better process efficiency. A hybrid DG system integrated with Compressed Air Energy Storage and Thermal Energy Storage is studied in [24]. Some scholars analyze the benefits of energy storage from an economic perspective. Authors in [25] propose a methodology for allocating an energy storage system in a distribution system with a high penetration of wind energy. The ultimate goal is to maximize the benefits for both the DG owner and the utility by sizing the storage to accommodate all amounts of spilled wind energy and by then allocating it within the system in order to minimize the annual cost of the electricity. On the other hand, two objectives are formulated with affine parameters including the minimization of total active power losses and the minimization of system voltage deviations in [26] to optimize the operation of storage in active distribution networks with uncertainties. Energy storage and DGs are planned in the distribution network simultaneously, which provides a more direct strategy for transforming the ordinary distribution network into ADNs.

In summary, we can find that the planning of DGs must take into account the fluctuation of their output, and energy storage has a good effect of smoothing the fluctuation of the power grid. If the planning of DGs is completed first and then the planning of energy storage devices is carried out according to the existing methods, they cannot cooperate and restrict each other to achieve the maximum economic benefits of distribution network. Based on the above reason, the DGs planning in this article sets two objectives: 1) the planning takes the annual cost as the objective function, and the cost is less than that of the existing methods. 2) The volatility of planning results is smaller and the system is more stable. Therefore, the joint planning method of DGs and storage is a better choice for the evolution from distribution network to active distribution network.

**1.2 Contribution of This Paper**

Table 1 summarizes the main differences between the proposed model in this paper and the most relevant research studies in the field. The main contributions of this paper are summarized as follows: 1) Energy storage and DGs are planned in the distribution network simultaneously, which provides a more direct strategy for transforming the ordinary distribution network into ADNs. 2) The application of the bi-level programming makes the location and capacity of DGs and energy storage interact with the operation of energy storage devices. In this way, the fluctuation of DGs is perfectly combined in the planning which makes the planning results more practical.

Table 1 Comparison of the proposed model with the most relevant studies

| Reference | Uncertainties | | Planning subject | | Main characteristics of the proposed model | The proposed methodology |
|---|---|---|---|---|---|---|
| | DGs | Load | DGs | Energy devices | | |
| 10 | × | × | √ | × | Single-level multi-objective model. The objective functions include voltage stability margin and power loss. | Dynamic programming search |
| 11 | × | √ | √ | × | Single-level multi-objective model. The objective functions include minimization of total active power losses, maximization of voltage stability index, minimization of total cost, and minimization of total emission produced by DGs and the grid. | Pareto-based multi-objective hybrid big bang-big Crunch algorithm |
| 12 | × | × | √ | √ | Multi-stage multi-objective framework. The objective functions include power loss, voltage stability, voltage deviation, installation cost, operational cost, and emission cost. | Hybrid metaheuristic algorithm |
| 18 | × | × | √ | √ | Single-level single-objective model. The objective function includes the investment cost, operation cost and they get profit by selling the renewable energy. | Grey wolf optimizer algorithm |
| 19 | × | × | √ | × | Single-level single-objective model. The objective is to maximize profit. | Mathematical method |
| 22 | √ | √ | √ | × | Single-level single-objective model. The objective function includes the investment cost, operating cost, maintenance cost, network loss cost, and capacity adequacy cost. | Monte Carlo simulation embedded genetic algorithm-based approach |
| This paper | √ | √ | √ | √ | Bi-level multi-objective model. The objective function of the upper-level is the minimum annual cost. The objective function of the lower-level is minimize the fluctuating operation cost | Improved binary particle swarm optimization algorithm based on chaos optimization |

### 1.3 Organization of This Paper

The remainder of this paper is organized as follows. Section 2 gives the uncertainty modelling of DGs. A detailed description of the joint planning model is put forward in Section 3. Section 4 demonstrates the model solution process in detail. Case studies on the PG & E 69-bus distribution system have been performed in Section 5. Finally, the conclusions are drawn in Section 6.

### 2. UNCERTAINTY MODELING of DGs

In this study, two kinds of renewable DGs, i.e. WT and PV, are studied. This section depicts the probabilistic models of these DGs in brief.

### 2.1 Probabilistic WT Model

The uncertainty of WT outputs is mainly originated from the inherent intermittency of wind speeds. Previous research demonstrates that wind speeds follow the Weibull distribution [27], [28]. The probability density function (PDF) of wind speeds is accordingly given by [29]

$$f_w(v) = (\iota/\gamma)(v/\gamma)^{k-1}\exp[-(v/\gamma)^{\iota}] \tag{1}$$

where $v$ represents the actual wind speed; $\iota$ is the shape factor (dimensionless), which describes the PDF shape of wind speeds; $\gamma$ is the scale factor.

The relationship between the WT power output $P^{WT}$ and the actual wind speed $v$ can be described as [28]:

$$P^{WT}(v) = \begin{cases} 0 & v < v_{in}, v > v_{out} \\ \dfrac{v - v_{in}}{v_* - v_{in}} P_* & v_{in} \leq v < v_* \\ P_* & v_* \leq v < v_{out} \end{cases} \quad (2)$$

where $P_*$ denotes the rated output power of a WT, $v_{in}$ is the cut-in wind speed, $v_{out}$ is the cut-out wind speed, and $v_*$ is the rated wind speed.

According to (1) and (2), the PDF of the WT output $f_o(P^{WT})$ can be formulated as

$$f_o(P^{WT}) = \begin{cases} (tHv_{in}/\gamma P_*)\left[((1+HP^{WT}/P_*)v_{in})/\gamma\right]^{t-1} \times \\ \quad \exp\left\{-\left[((1+HP^{WT}/P_*)v_{in})/\gamma\right]^t\right\}, p^{WT} \in [0, P_*] \\ 0, \quad \text{otherwise} \end{cases} \quad (3)$$

where $H = (v_*/v_{in}) - 1$.

## 2.2 Probabilistic PV Model

The PV output is mainly dependent on the amount of solar irradiance reaching the ground, ambient temperature and characteristics of the PV module itself. The statistical study shows that the solar irradiance for each hour of the day follows the Beta distribution [30], which is a set of continuous probability distribution functions defined in interval (0, 1). The Beta PDF used to depict the probabilistic nature of solar irradiance is given by

$$f_r(r) = \frac{\Gamma(\lambda_1) + \Gamma(\lambda_2)}{\Gamma(\lambda_1)\Gamma(\lambda_2)} \left(\frac{r}{r_{max}}\right)^{\lambda_1 - 1} \left(1 - \frac{r}{r_{max}}\right)^{\lambda_2 - 1} \quad (4)$$

where $r$ and $r_{max}$ are respectively the actual light intensity and its maximum value; $\lambda_1$ and $\lambda_2$ are the shape factors, $\lambda_1 = \mu_{pv}\left(\dfrac{\mu_{pv}(1-\mu_{pv})}{\sigma_{pv}^2} - 1\right)$, $\lambda_2 = (1-\mu_{pv})\left(\dfrac{\mu_{pv}(1-\mu_{pv})}{\sigma_{pv}^2} - 1\right)$; $\Gamma$ represents a Gamma function in the following form: $\Gamma(\lambda) = \int_0^{+\infty} \rho^{\lambda-1} e^{-\rho} d\rho$, wherein $\rho$ is an integer variable. The relationship between PV outputs and solar irradiances is [27].

$$P^{PV} = \xi \eta_m A_{pv} \eta_{pv} \cos\theta \quad (5)$$

where $\xi$ is the solar irradiance, $\eta_m$ is the maximum power point tracking, $A_{pv}$ is the radiation area of this PV, $\eta_{pv}$ is the conversion efficiency, and $\theta$ is the solar incident angle.

From (5) it can be seen that the PV output is linear with the solar irradiance, and thereby, the PV output is also generally subject to the Beta distribution. The PDF of PV output is [27].

$$f_p(P^{PV}) = \frac{\Gamma(\lambda_1) + \Gamma(\lambda_2)}{\Gamma(\lambda_1)\Gamma(\lambda_2)} \left(\frac{P^{PV}}{P_{max}^{PV}}\right)^{\lambda_1 - 1} \left(1 - \frac{P^{PV}}{P_{max}^{PV}}\right)^{\lambda_2 - 1} \quad (6)$$

where $P^{PV}$ and $P_{max}^{PV}$ represent the output of this PV and its maximum value, respectively.

## 2.3 SERIALIZATION MODELING OF RANDOM VARIABLES
### 2.3.1 Introduction of Discretized Step Transformation

The sequence operation theory (SOT) is a powerful mathematical tool to handle multiple uncertainties, which is based on the sequence convolution in the field of digital signal processing

[31]. The key idea of SOT is based on the concept of sequence operations: first, continuous random variables are discretized as probabilistic sequences according to a given discrete step by using their respective PDFs, and then a newly generated sequence is obtained via mutual operations.

**Definition 1.** Suppose a discrete sequence $a(i)$ with the length $N_a$, $a(i)$ is called a probabilistic sequence if

$$\sum_{i=0}^{N_a} a(i) = 1, \ a(i) \geq 0, \ i = 0,1,2,...,N_a \tag{7}$$

**Definition 2.** Given a probabilistic sequence $a(i)$ with the length $N_a$, its expected value is defined as follows:

$$E(a) = \sum_{i=0}^{N_a}[i \ a(i)] = \sum_{i=1}^{N_a}[i \ a(i)] \tag{8}$$

Two kinds of sequence operations, i.e. addition-type-convolution (ATC) and subtraction-type-convolution (STC), are defined as follows.

**Definition 3.** Given two discrete sequences $a(i_a)$ and $b(i_b)$, with length $N_a$ and $N_b$. The ATC and STC are defined as:

$$gs_1(i) = \sum_{i_a + i_b = i} a(i_a) b(i_b), \quad i = 0,1,2,...,N_a + N_b \tag{9}$$

$$gs_2(i) = \begin{cases} \sum_{i_a - i_b = i} a(i_a) b(i_b), & 1 \leq i \leq N_a \\ \sum_{i_a \leq i_b} a(i_a) b(i_b), & i = 0 \end{cases} \tag{10}$$

where $gs_1(i)$ and $gs_2(i)$ are called generated sequences.

**2.3.2 Sequence Description of Intermittent DG Outputs**

Taking WT as an example, the sequence description of DG outputs are described below. During a time period $t$, the WT output $P_t^{WT}$, PV output $P_t^{PV}$, and load power $P_t^L$ are all random variables, and they can be depicted by the corresponding probabilistic sequences $a(i_{at})$, $b(i_{bt})$ and $d(i_{dt})$ through discretization of continuous probability distributions. The length of WT output probabilistic sequence $N_{at}$ is calculated by

$$N_{at} = [P_{\max,t}^{WT} / q] \tag{11}$$

where $q$ denotes the discrete step size and $P_{\max,t}^{WT}$ is the maximum value of WT power output during period $t$.

Table 2 shows the WT output and its corresponding probabilistic sequences.

Table 2 WT output and its probabilistic sequence

| Power (kW)  | 0    | q    | … | $u_a q$  | … | $N_a q$  |
|-------------|------|------|---|----------|---|----------|
| Probability | a(0) | a(1) | … | $a(u_a)$ | … | $a(N_a)$ |

The probabilistic sequence of the WT output can be calculated by using its PDF, which is given as follows:

$$a(i_{at}) = \begin{cases} \int_0^{q/2} f_o(P^{WT}) dP^{WT}, & i_{at} = 0 \\ \int_{i_{at}q - q/2}^{i_{at}q + q/2} f_o(P^{WT}) dP^{WT}, & i_{at} > 0, i_{at} \neq N_{at} \\ \int_{i_{at}q - q/2}^{i_{at}q} f_o(P^{WT}) dP^{WT}, & i_{at} = N_{at} \end{cases} \tag{12}$$

In this paper, a 24-hour time interval is set to the planning work, and it is convenient to use the sequence operation theory to get the expectation value of the stochastic outputs of WT and photovoltaic power generation in an hour. After analyzing the average wind speed and light intensity of each time interval in a year, the probability density distribution of output of WT and photovoltaic power generation is calculated, and the theory of sequence operation is applied to discretize and serialize the data, thus the hourly expectation value of the DGs can be obtained.

The probability sequence of the output of the wind power generator at the $t$-th sampling is set to $a(i_{a,t})$, and its sequence length is $N_{a,t}$. The probabilistic sequence of PV output at the $t$-th sampling is set to $b(i_{b,t})$, and its sequence length is $N_{b,t}$. Then the expected outputs of the two kinds of intermittent DGs are $E_t^{\text{WT}}$ and $E_t^{\text{PV}}$:

$$E_t^{\text{WT}} = \sum_{m_{at}=0}^{N_{at}} m_{at} q \cdot a(m_{at}) \tag{13}$$

$$E_t^{\text{PV}} = \sum_{m_{bt}=0}^{N_{bt}} m_{bt} q \cdot b(m_{bt}) \tag{14}$$

It should be noted that the outputs of PV units might be zero without light, and in this case WT units provide the total outputs of DGs.

## 3. Joint planning model

Multi-level modeling can simplify problems when solving many similar problems [32, 33]. In the joint optimal configuration model of this paper, the installation position and capacity of DGs and energy storage devices are optimized with the minimum economic objective function in the first planning part, but the network power loss and the intra-day scheduling of energy storage devices cannot be obtained until the intra-day optimization is finished. On the other hand, the optimization of the intra-day scheduling of energy storage devices in the second part is based on the installation position and capacity of DGs and the energy storage devices. Considering the inherent hierarchical features between the two parts of the model, a bi-level programming model is proposed in this study.

### 3.1 Bi-level programming theory

The bi-level model is constructed to solve the optimization problem of a bi-level hierarchical structure system. Both the upper-level model and the lower-level model have their own objective functions and constraints. The basic process is that the upper-level model gives the lower-level model a variable, then the lower-level model takes this variable as the decision variable and calculates the objective function under the constraint conditions. After obtaining the optimal value of the objective function in the lower-level model, the optimal value is fed back to the upper-level model to calculate its objective function under the constraint conditions [34]. The general formulas of the nonlinear bi-level model are:

$$\text{Min} \quad F = F(x,v) \tag{15}$$

$$\text{s.t} \quad G(x) \leq 0 \tag{16}$$

$$\text{Min} \quad v = f(x,y) \tag{17}$$

$$\text{s.t} \quad g(x,y) \leq 0 \tag{18}$$

where equation (15) is the objective function of upper-level model, and $x$ and $v$ are its decision variables; equation (16) is the constraint of upper-level model; equation (17) is the objective function of lower-level model, and $x$ and $y$ are its decision variables; equation (18) is the constraint of the lower-level model. The objective function and constraints of lower-level model are affected when the decision variable *x* of the upper-level model is passed to the lower-level model; similarly, the optimal value of the objective function of the lower-level model *v* is passed to the upper-level model as the decision variable, affecting the objective function of the upper-level model. Overall, the interaction between the upper and lower-level model is fulfilled.

The bi-level programming model proposed in this paper is as follows: The upper-level model optimizes the position and capacity of energy storage and DGs to find the optimal annual economic objective function, and then transfers the position and capacity of energy storage and DGs to the lower-level model. The lower-level model optimizes the intraday scheduling of energy storage in every season to obtain the optimal annual fluctuating operation cost, and then return the annual fluctuating operation cost to the upper-level model to generate the annual economic objective function. The location and capacity of the energy storage and DGs in the upper-level model are equivalent to "*x*" in the bi-level model, and the annual fluctuating operation cost in the lower-level model is equivalent to "*v*" in the bi-level model, which makes the iteration between the two levels operates. The structure of the proposed model is shown in Fig. 1.

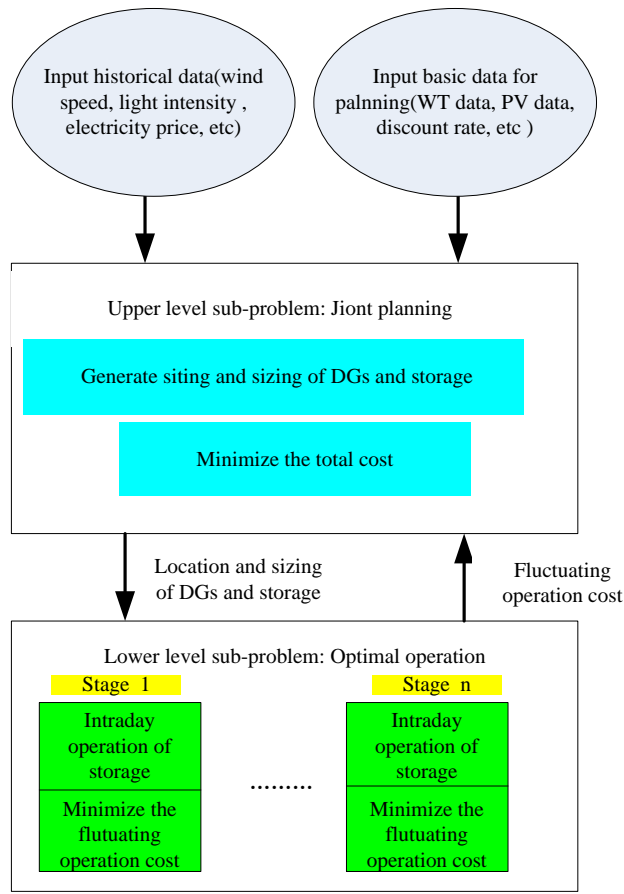

Fig. 1 The structure of the proposed model

## 3.2 Upper-level model

(1) Objective function

In this study, the upper-level aims to seek the optimal location and capacity of DGs and energy storage, and the objective function of the upper-level model is the minimum annual cost. The annual cost includes annual equipment investment cost $C_1$, annual operation and maintenance cost $C_2$, and annual power purchase cost from the power grid $C_3$.

The objective function of the upper-level model is expressed as:

$$\min F_1 \tag{19}$$

$$F_1 = C_1 + C_2 + C_3 \tag{20}$$

1) This paper mainly studies the planning of DGs and energy storage devices, so the annual equipment investment cost $C_1$ can be given by:

$$C_1 = C_f C_{wd} S_{wd} + C_g C_{pv} S_{pv} + C_E (C_{ST}^{INSE} E^{ST} + C_{ST}^{INSS} S^{ST}) \tag{21}$$

where $C_f$, $C_g$ and $C_E$ are respectively the present coefficients of WT, PV and energy storage devices, which are related to the discount rate and service life of the devices; $C_{wd}$ and $C_{pv}$ are the investment costs of WT and PV per kW respectively; $S_{wd}$ and $S_{pv}$ are the total capacities of WT and PV respectively; $C_{ST}^{INSE}$ and $C_{ST}^{INSS}$ are respectively the investment cost of energy storage devices related to rated power and installation capacity; $E^{ST}$ is the rated power of energy storage devices; $S^{ST}$ is the installation capacity of energy storage devices.

2) The operation of the DG units and inverters needs to be managed and maintained. The maintenance cost is generally dependent on the performance, service life and the price of the fittings, the amount of its expenditure is relatively fixed. The operation cost is related to the rated power of DGs, that is, the operation cost is proportional to the output power of DGs. The operation and maintenance cost of energy storage devices is related to its rated power.

In the model, it is divided into four seasons each year, and each season uses a representative day to characterize the outputs of DGs. Firstly, the field data of a certain place in three years are collated, and then these data are used to generate the probability density distributions of average wind speed and light intensity; secondly, each day (a scheduling cycle) is divided into 24 time periods, each period is takes as an hour, and each hour is adopted as the representative hour of each season, so that there are 96 representative hours of probability density distributions of average wind speed and light intensity; finally, the expected outputs of each hour is obtained by the sequence operation theory. To facilitate the calculation of the annual cost, the number of days per season is taken as 91 in this work.

Here, the total operation and maintenance cost $C_2$ is

$$C_2 = (C_{wd} S_{wd} + C_{pv} S_{pv}) y + \sum_{s=1}^{4} 91 \sum_{h=1}^{24} [z(P_{wd,sh} + P_{pv,sh})] + C_{ST}^{OM} S^{ST} \tag{22}$$

where $y$ is the maintenance cost coefficient of WT and PV related to the rated power; $P_{wd,sh}$ and $P_{pv,sh}$ are respectively the expected outputs of WT and PV of hour $h$ in season $s$; $z$ is the DGs' operation cost related to the output power; $C_{ST}^{OM}$ is the operation and maintenance cost of energy storage devices per year related to the rated power.

3) The power purchase cost from the power grid is also part of the annual fee. The power purchase cost from the upstreaming power grid $C_3$ is shown as follows

$$C_3 = \sum_{s=1}^{4} 91 \sum_{h=1}^{24} [C_{pu,sh}(P - P_{wd,sh} - P_{pv,sh} + P_{loss,sh} - P_{dc,sh} + P_{ch,sh})] \tag{23}$$

where $P$ is the active load of the system; $C_{pu,sh}$ is the time-of-use electricity price of $h$-th hour in the $s$-th season; $P_{loss,sh}$ is the total system active power losses of $h$-th hour in the $s$-th season; $P_{ch,sh}$ and $P_{dc,sh}$ are the charge power and discharge power of energy storage devices of $h$-th hour in the $s$-th season, respectively.

(2) Technical constraints

The maximum capacity of DGs is limited by the following conditions:

$$\sum_{i=2}^{N} P_{DG,i} \leq \sum_{i=2}^{N} P_{D,i} + P_{LO} \qquad (24)$$

$$\sum_{i=2}^{N} Q_{DG,i} \leq \sum_{i=2}^{N} Q_{D,i} + Q_{LO} \qquad (25)$$

where $P_{DG,i}$ and $Q_{DG,i}$ are the active power and reactive power of DG outputs at the bus $i$ respectively; $P_{D,i}$ and $Q_{D,i}$ are the active power and reactive power of load at the bus $i$ respectively; $P_{LO}$ and $Q_{LO}$ are the total active power losses and reactive power losses of the system without DGs.

### 3.3 Lower-level model

The lower-level model takes the daily operation of the distribution network as the optimization scenario, in which the location and capacity of DGs are determined by the upper-level model, and the outputs of DGs in an hour are determined by sequence operation theory. In this scenario, the charge-discharge power of energy storage devices is optimized in the direction of the optimal economy.

(1) Objective function

$$\min F_2 \qquad (26)$$

Influenced by the time-of-use electricity price, the daily charge-discharge power of energy storage devices and the fluctuating active power loss will affect the power purchase cost. The part of the power purchase cost determined by active power loss and the charge-discharge power of energy storage devices for 24 hours in the $s$-th season, which belongs to parts of $C_3$ in the upper-level model is as follows:

$$F_2 = \sum_{h=1}^{24} C_{pu,h}(P_{loss,h} + P_{e,h}) \qquad (27)$$

where $C_{pu,h}$ is the time-of-use electricity price of $h$-th representative hour in a day; $P_{loss,h}$ is the active power loss of $h$-th representative hour in a day; $P_{e,h}$ is the charge-discharge power of energy storage devices of $h$-th representative hour in a day.

(2) Technical constraints

Due to an ADN has a radial topology and allows for bidirectional power flow, this work uses the forward-backward sweep method to solve the power flow. The branch flow equations are used to describe the power flow and are defined as follows:

$$P_{h,i} = \sum_{j=1, j \neq i}^{N} Y_{ij} U_{h,i} U_{h,j} \cos(\theta_{ij} + \delta_j - \delta_i) \qquad (28)$$

$$Q_{h,i} = \sum_{j=1, j \neq i}^{N} Y_{ij} U_{h,i} U_{h,j} \sin(\theta_{ij} + \delta_j - \delta_i) \qquad (29)$$

$$U_{h,i}^2 - U_{h,j}^2 - 2(r_{ij} P_{h,ij} + x_{ij} Q_{h,ij}) + (r_{ij}^2 + x_{ij}^2) I_{h,ij}^2 = 0 \qquad (30)$$

$$I_{h,ij}^2 U_{h,i}^2 = P_{h,ij}^2 + Q_{h,ij}^2 \qquad (31)$$

$$U_{min}^2 \leq U_{h,i}^2 \leq U_{max}^2 \qquad (32)$$

$$I_{h,ij}^2 \leq I_{max}^2 \qquad (33)$$

The AC power flow equations are denoted by (28) to (31), where these equations represent the active power balance, reactive power balance, the voltage drop constraint and complex power flows of the lines, respectively. In addition, the system voltage limits and the maximum line current capacity are represented in (32) to (33).

The total power consumption should be equal to the total power supply at each bus:

$$P_{Swing} + \sum_{i=2}^{N} P_{DG,i} = \sum_{i=2}^{N} P_{D,i} + P_L \tag{34}$$

$$Q_{Swing} + \sum_{i=2}^{N} Q_{DG,i} = \sum_{i=2}^{N} Q_{D,i} + Q_L \tag{35}$$

Energy storage devices must meet the following constraints:

$$\begin{cases} 0 \leq P_{ch,h} \leq P_{ch,\max} \\ 0 \leq P_{dc,h} \leq P_{dc,\max} \end{cases} \tag{36}$$

$$S_{oc\min} \leq S_{oc,h} \leq S_{oc\max} \tag{37}$$

$$S_{oc,h} = S_{oc,h-1} + (\eta_{ch} P_{ch,h} - P_{dc,h}/\eta_{dc})\Delta t \tag{38}$$

$$S_{oc,0} = S_{oc,24} \tag{39}$$

## 4 Model solution

For addressing the proposed planning model, this paper adopts improved binary particle swarm optimization algorithm based on chaos optimization (IBPSO) to solve the upper- and lower- level models respectively. Through the iteration within the algorithm and in the bi-level model, the global optimal solution in the planning and operation is transmitted to each other, and the optimal allocation of DGs is finally obtained.

### 4.1 Binary particle swarm optimization algorithm

The original BPSO was proposed by Kennedy and Eberhart [35] to allow PSO to operate in binary problem spaces. In this version, particles could only fly in a binary search space by taking values of 0 or 1 for their position vectors. The roles of velocities are to present the probability of a bit taking the value 0 or 1. The velocities are mathematically modeled as follow [36]:

$$v_n^k(iter+1) = w \cdot v_n^k(iter) + c_1 \cdot rand \cdot \left[ pbest_n - x_n^k(iter) \right] + c_2 \cdot rand \cdot \left[ gbest_n - x_n^k(iter) \right] \tag{40}$$

where $v_n^k(iter)$ is the velocity of particle $n$ at iteration $iter$ in $k$-th dimension, $w$ is a weighting function, $c_1$ and $c_2$ are acceleration coefficients, $rand$ is a random number in the range [0,1], $x_n^k(iter)$ is the current position of particle $n$ at iteration $iter$ in the $k$-th dimension, $pbest_n$ is the best solution that the $n$-th particle has obtained so far, and $gbest_n$ indicates the best solution the swarm has obtained so far.

A sigmoid function (transfer function part) as in Eq. (41) was employed to transform all real values of velocities to probability values in the interval [0,1].

$$T\left(v_n^k(iter)\right) = \frac{1}{1+e^{-v_n^k(iter)}} \tag{41}$$

After converting velocities to probability values, position vectors could be updated with the probability of their velocities as follow:

$$x_n^k(iter+1) = \begin{cases} 0 & \text{If } rand < T\left(v_n^k(iter+1)\right) \\ 1 & \text{If } rand \geq T\left(v_n^k(iter+1)\right) \end{cases} \tag{42}$$

## 4.2 Proposed IBPSO algorithm

As the "No free lunch" theorem points out, there is no existing such an algorithm that works best for all performance indices like global optimization ability and convergence speed. As far as the original BPSO is concerned, BPSO allows PSO to operate in binary problem spaces, but it also has the problem of individuals prematurely converge to a local optimum. Taking into account the inherent natures of chaotic motion like ergodicity, randomness and high sensitivity to initial conditions, a Tent-map-based chaotic search strategy is put forward for handling the premature.

Recent research has demonstrated that the population fitness variance (PFV) $\sigma^2$ is an effective indicator being used for dynamically monitoring the degree of population crowding during the evolutionary process [37, 38]. The indicator is

$$\sigma^2 = \sum_{m=1}^{N_p} \left( \frac{f_m - f_{avg}}{f_{best}} \right)^2 \quad (43)$$

where $f_m$ is the fitness value of bacterial $m$; $f_{avg}$ and $f_{best}$ are the average value and best fitness of the population; $N_p$ is the population size.

As is known, it is a critical problem when studying premature convergence to identify its occurrence and extent [39]. For addressing this issue, we develop a novel dynamic monitoring mechanism by measuring the population fitness variance during the optimization process. Once premature convergence occurs, optimization variables will be mapped into chaotic variables through carrier waveforms, and the chaos optimization will be immediately started up to maintain the population diversity and avoid premature convergence. The criterion for detection of premature convergence is

$$low\_thr < \frac{\sigma^2_{iter+1}}{\sigma^2_{iter}} < up\_thr \quad (44)$$

Here, $\sigma^2_{iter+1}$ and $\sigma^2_{iter}$ are the values of PFV in iteration $iter+1$ and $iter$, respectively. $low\_thr$ and $up\_thr$ are respectively taken as 0.99 and 1.01 since they give the most satisfactory results through large amounts of tests.

Considering the traversal inhomogeneity of the Logistic map, the Tent map is chosen as the chaos map as follows.

$$t_{iter+1} = \begin{cases} 2t_{iter}, & 0 \leq t_{iter} \leq 1/2 \\ 2(1-t_{iter}), & 1/2 \leq t_{iter} \leq 1 \end{cases} \quad (45)$$

When the periodic points (0.2, 0.4, 0.6, 0.8) or the fixed points (0, 0.25, 0.5, 0.75) occurs in the iterative sequence of the Tent map, the current variable $t_{iter+1}$ is updated by adding random perturbations timely to make the map re-enter the chaotic state, as shown in (46).

$$t_{iter+1} = \frac{t_{iter+1} + \text{rand}(0,1)}{2} \quad (46)$$

The flowchart of the proposed IBPSO algorithm is show as follows:

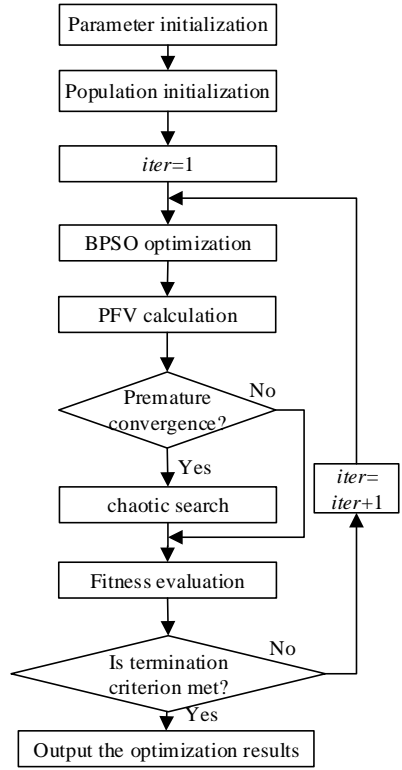

Fig. 2 Flowchart of the proposed IBPSO algorithm

### 4.3 Solving process

The solving process of the proposed bi-level planning model mainly includes the following steps:

*Step 1*: Input the initial variables, including system parameters, variable range, and the IBPSO parameters.

*Step 2*: Initialize the particle swarm of the upper-level model, including the positions and velocities of the particles.

*Step 3*: The positions of particles, i.e. the location and capacity of DGs and energy storage devices, are passed to the lower-level model. According to the installed capacity of DGs given by the upper-level model, the 24-hour power outputs of these generators in four seasons are obtained, and then the lower-level model is initialized.

*Step 4*: The lower-level model optimizes the intraday scheduling of energy storage to get the minimum fluctuating operation cost by using the IBPSO, and then changes the output of DGs of four seasons to form the annual cost. Accordingly, the optimal annual fluctuating operation cost is obtained.

*Step 5*: Calculate the objective function of the lower-level model for all particles, then return the optimal results of fluctuating operation cost to the upper-level model.

*Step 6*: Calculate the objective function of the upper-level model for all particles and check the technical constraints.

*Step 7*: Update all the particles' velocities and positions of the upper-level model to form a new particle swarm.

*Step 8*: Check whether the termination criterion is met. If the maximum number of iterations is reached, terminate the optimization process and output the optimal result; otherwise, the iteration counter *iter = iter*+1, and go to *Step 3*.

# 5 Case studies
## 5.1 Parameter settings

In this paper, the PG & E 69-bus system is used for simulation analysis. Fig. 3 shows the network framework of the PG & E 69-bus system, whose voltage level is 12.66 kV, the total active load is 3715 kW and the total reactive load is 2300 kVar. The parameters of the system circuit are listed in the literature [20]. The parameters of IBPSO are as follows: the number of particles is 50; the maximum number of iterations is 100; the maximum value of the inertia weight is 0.9 and the minimum value is 0.4.

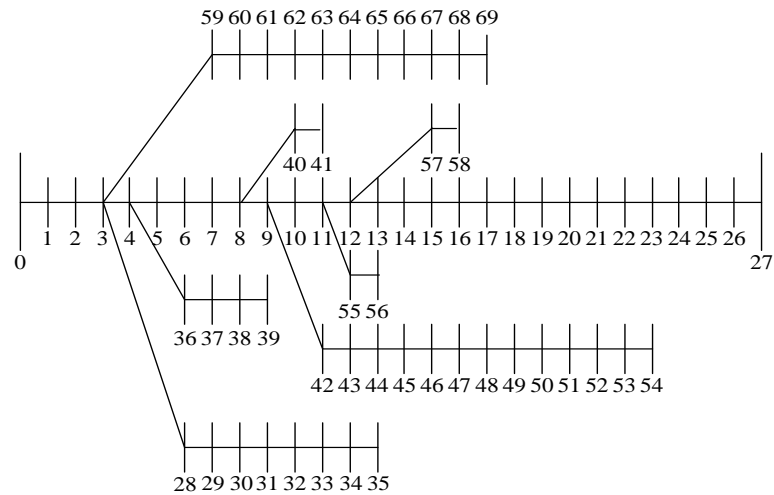

Fig. 3 The PG & E 69-bus system

The three-stage electricity price based on the time period is the system's time-of-use electricity price, in which the electricity price in spring and summer are the same, and the electricity price in autumn and winter are the same [40]. The specific electricity price is shown in Fig. 4.

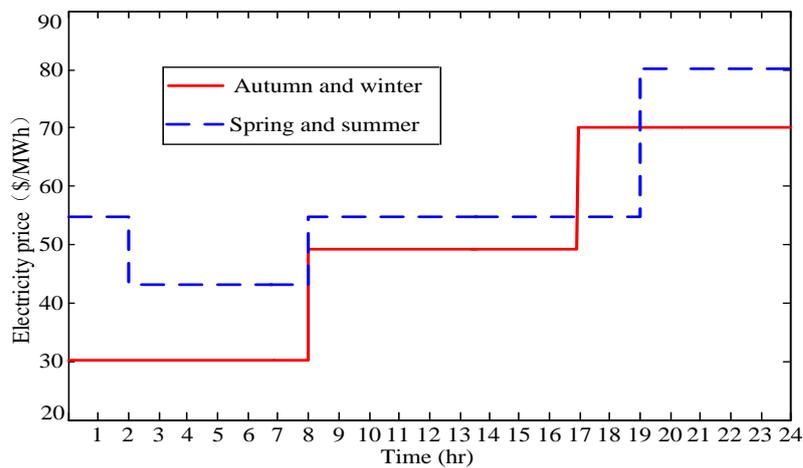

Fig. 4 Time-of-use electricity price at various times throughout the year

Employ the network sensitivity location index which is widely used, this paper select bus 49, 50, 61 and 64 as candidate installation buses of DGs for the purpose of reducing network loss. The economic parameters of WT and PV are shown in Table 3. Considering the reduction of network loss, installation buses of energy storage are the same as DGs, and zinc/bromine flow battery which is the most economical is adopted. The parameters are shown in Table 4.

Table 3 Economic parameters of WT and PV

| Type of DG | Investment cost ($/kWh) | Operating cost ($/kWh) | Maintenance cost coefficient ($/kWh) | Present value coefficient |
|---|---|---|---|---|
| WT | 1230 | 0.015 | 0.015 | 0.0802 |
| PV | 1540 | 0.015 | 0.015 | 0.071 |

Table 4 Economic parameters of zinc / bromine battery

| $C_{ST}^{INSE}$ ($/kWh) | $C_{ST}^{INSS}$ ($/kW) | $C_{ST}^{OM}$ ($/kW) | $C_E$ (pu) |
|---|---|---|---|
| 232 | 180 | 21 | 0.037 |

## 5.2 Results and discussions
### 5.2.1 Comparison of different DG configuration types

The economic analysis results of the bi-level planning are performed in the following four scenarios, in which the maximum capacity of DGs is limited to 30% of the total load and the energy storage device is limited to 10%.

Scenario 1 is a traditional distribution network model without considering the integration of WT, PV and Energy Storage;

Scenario 2 is the joint Planning of WT, PV and Energy Storage;

Scenario 3 is the joint Planning of WT and Energy Storage;

Scenario 4 is the joint Planning of PV and Energy Storage.

Table 5 Comparison of optimal allocation results

| Scenarios | Wind turbine /kW （bus） | Photovoltaic unit /kW （bus） | Energy storage device /kWh (bus) |
|---|---|---|---|
| Scenario 1 | — | — | — |
| Scenario 2 | 782(61) | 59(49)  48(50) | 370(61) |
| Scenario 3 | 826(61)  83(49)  38(64) | — | 282(61) |
| Scenario 4 | — | 680(61)  142(49) | 323(61) |

Table 6 Comparison of the annual cost of optimal allocation (M$)

| Scenarios | Total cost | Investment cost | Operation and maintenance costs | Charge for electricity purchase |
|---|---|---|---|---|
| Scenario 1 | 2.71 | — | — | 2.71 |
| Scenario 2 | 1.94 | 0.11 | 0.21 | 1.62 |
| Scenario 3 | 1.93 | 0.13 | 0.24 | 1.56 |
| Scenario 4 | 1.99 | 0.12 | 0.14 | 1.73 |

Analyzing the planning results of four scenarios in Table 5 and Table 6, the following conclusions can be drawn:

(1) The electricity price in Scenario 1 is set to 50 $/MWh , and the total power consumption of the system is the sum of total load and network loss. From Table 4, we can see that the total cost of Scenario 1 is higher than that of the other three scenarios, which shows that the distributed

generation planning considering the time-of-use price is very effective for economic benefit.

(2) Scenario 3 has the largest DGs capacity, the smallest storage capacity, and the least cost. From the comparison between Scenario 3 and Scenario 2, 4, it can be found that Scenario 3 has the largest installed WT capacity and the least cost, while Scenario 4, where all the units are PV, has the most cost. It can be seen that the economic benefits created by WT are higher, partly because the investment and operating costs of wind turbines are lower, and another part is that WT have a long effective working time throughout the year, and PV can only work during periods the of light. Therefore, the optimized result is in line with our objective function, that is, the economic benefit is maximized. At the same time, PV have higher demand for energy storage devices, which can be reflected in the installation capacity of energy storage in Scenario 3 and Scenario 4.

Through the comparison of the above four scenarios, it can be seen that the access of distributed generation sources in the distribution network effectively improves the economic benefits of the distribution network, and the benefit of wind turbines is better. But in actual planning, affected by the geographical environment, different regions may be suitable for installing different types of DGs, so planners need to find a compromise between environmental and economic factors, finally achieve the optimum profit through the joint planning method in this paper.

**5.2.2 Result analysis of time period operation planning**

In the planning model, the 24-hour operating state of the system is also considered. According to the lower-level model, the output of WT, PV and storage and the active power loss are listed. The operation of the specific period in winter is shown in Table 7, where charge is positive and discharge is negative.

Table 7 Intraday operation in winter

| Time interval | WT output /kW (bus 61) | PV output /kW (bus 49) | PV output /kW (bus 50) | Charging/discharging of energy storage /kW | Active power loss /kW |
|---|---|---|---|---|---|
| 0 | 381.5 | 0 | 0 | 0 | 171.4 |
| 1 | 396.76 | 0 | 0 | 0 | 169.5 |
| 2 | 366.24 | 0 | 0 | 40 | 178.4 |
| 3 | 381.5 | 0 | 0 | 0 | 171.4 |
| 4 | 396.76 | 0 | 0 | 30 | 173.2 |
| 5 | 404.39 | 0 | 0 | 0 | 168.5 |
| 6 | 419.65 | 0 | 0 | 0 | 166.6 |
| 7 | 457.8 | 5.18 | 4.8 | 0 | 162.1 |
| 8 | 495.95 | 6.22 | 5.76 | 0 | 157.7 |
| 9 | 518.84 | 7.25 | 6.72 | 0 | 155.1 |
| 10 | 534.1 | 7.77 | 7.2 | 40 | 157.9 |
| 11 | 572.25 | 9.06 | 8.4 | 40 | 153.6 |
| 12 | 595.14 | 10.36 | 9.6 | 0 | 146.9 |
| 13 | 534.1 | 9.06 | 8.4 | 0 | 153.4 |
| 14 | 549.36 | 8.03 | 7.44 | 0 | 151.8 |
| 15 | 518.84 | 6.48 | 6 | 0 | 155.1 |
| 16 | 495.95 | 5.18 | 4.8 | 0 | 157.8 |
| 17 | 457.8 | 0 | 0 | 0 | 162.2 |
| 18 | 427.28 | 0 | 0 | -20 | 163.4 |
| 19 | 457.8 | 0 | 0 | -30 | 158.7 |
| 20 | 473.06 | 0 | 0 | -20 | 158.1 |
| 21 | 442.54 | 0 | 0 | -30 | 160.4 |
| 22 | 419.65 | 0 | 0 | -30 | 163.1 |
| 23 | 396.76 | 0 | 0 | -20 | 167 |

From Table 5, it can be observed the following facts:

(1) Wind turbines are available throughout the day and peak at 8-16 pm, while photovoltaic units are only available at 7-16 pm. 8:00-16:00 is the common period of wind turbine output peak and photovoltaic unit output when the active power network loss of the system is lower than other periods. This shows that the distributed generation sources can effectively reduce the network loss after connecting to the distribution network, and the active power loss also decreases with the increase of the output of DGs.

(2) From the results of the optimization of the daily charge and discharge of the energy storage device, it can be seen that storage is charged or keep still in period 0:00-17:00 and discharged in period 18:00-23:00. This shows that the energy storage device, as a controllable power supply, charges in the period of low electricity price and discharges in the period of high electricity price, which conforms to the optimization direction of the lower economic objective function. At the same time, it can be seen in period 18:00-23:00, when WT output is low and PV output is zero, the energy storage device is charged to plays a role in mitigating the fluctuation caused by DGs.

### 5.2.3 Comparison of different algorithms and models

In order to verify the applicability of the IBPSO used in the article, the authors use Tabu search and BPSO to replace the IBPSO in this paper. The calculation results and speed are shown in the following Table 8.

Table 8 Comparison of different algorithms

| Algorithm name | Total cost (M$) | Duration of calculation(hour) |
|---|---|---|
| IBPSO (This paper) | 1.94 | 0.86 |
| Tabu search | 1.96 | 0.92 |
| BPSO | 1.95 | 0.96 |

It can be seen from Table 8 that the total cost of Tabu search and BPSO are higher than the IBPSO in this paper, and the IBPSO has decreased the duration of calculation by 6~10% compared to other two algorithms, so the IBPSO is adopted in this paper.

To show the useful property of bi-level optimization method, the outcomes of the centralized version of the model is analyzed. The centralized version of the model is divided into two parts, firstly complete the planning of the capacity and location of DGs, and secondly planning the capacity and location of storage. The results of the two methods are compared in the case of Scenario 2, which are shown in the following Table 9.

Table 9 Comparison of different models (M$)

| Model name | Total cost | Investment cost | Operation and maintenance costs | Charge for electricity purchase |
|---|---|---|---|---|
| Bi-level optimization model | 1.94 | 0.11 | 0.21 | 1.62 |
| Centralized version model | 2.05 | 0.12 | 0.22 | 1.71 |

As can be seen from Table 9 that the bi-level optimization method is better than the centralized version of the model in terms of both the cost of each part and the total cost.

**5.2.4 Impact analysis of energy storage access capacity**

In order to analyze the impact of energy storage, the cost indices of the proposed joint planning model and the model that only considers the DGs planning are shown in Fig. 5.

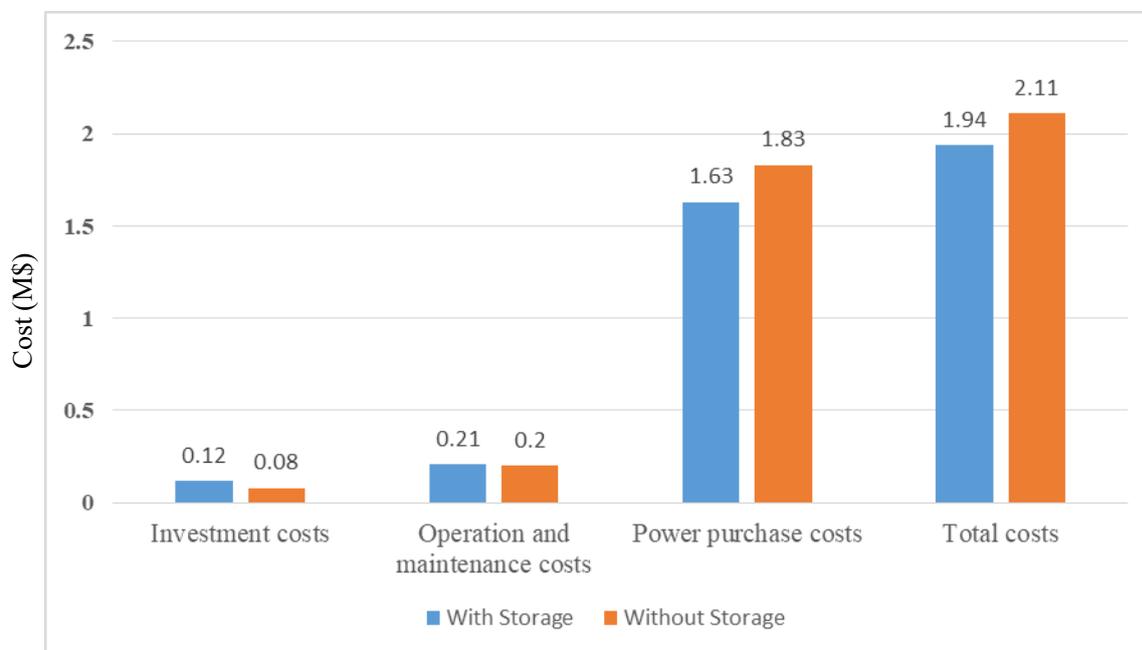

Fig. 5 Impact analysis of energy storage planning

It can be seen from Fig. 5 that energy storage devices have a significant influence. In the model that considering energy storage device planning in this paper, the cost of investment and operation and maintenance increases, but the cost of electricity purchase decreases greatly. On the one hand, the charging and discharging of storage with time-of-use electricity price brings incomes; on the other hand, the storage's peak shaving effect on the reduction of active power loss effectively improve the economic benefits.

To further analyze the impact of energy storage on distribution network planning, Table 10 lists the impact of access capacity of storage. This table shows the objective function of the proposed planning model under the conditions of different energy storage access capacity limitation from 10% to 30%.

Table 10 Impact analysis of energy storage access capacity limitation

| Storage penetration (%) | Storage capacity /kW (bus) | Investment cost (M$) | Operation and maintenance costs (M$) | Charge for electricity purchase (M$) | Total cost (M$) |
|---|---|---|---|---|---|
| 10 | 370(61) | 0.12 | 0.21 | 1.63 | 1.94 |
| 20 | 735（61） | 0.15 | 0.22 | 1.55 | 1.92 |
| 30 | 1110（61） | 0.18 | 0.23 | 1.52 | 1.93 |

As can be seen from the table, when the storage capacity increases, the corresponding investment, operation, and maintenance costs increase, while the purchase cost decreases. This shows that the energy storage device carries out an economical charging and discharging strategy according to the time-of-use electricity price in the planning, and its peak shaving ability also effectively reduces the network loss, which leads to the reduction of electricity purchase costs. It is noteworthy that when the storage access capacity limitation rises from 20% to 30%, the purchase cost decreases little, while the total cost increases slightly, which indicates that when the storage capacity increases continuously, its contribution to reducing cost is close to saturation. Of course, there are some models in which more than 100% of the storage capacity also has a very high economic benefit [35], mainly because they use objective function that also consider stability, including reducing outage time and providing islanding capacity, which is not considered in this paper.

**5.2.5 Analysis of the influence of energy storage to network voltage**

Different from the separate planning model of DGs, joint planning model considering both DGs and energy storage in this paper performs better in improving the system voltage. In the case studies of the paper, the voltage amplitude of bus 27 is the lowest and has the largest fluctuation in the above-mentioned scenarios. Fig. 6 shows the voltage amplitude of joint planning model and separate planning model of DGs. The cumulative probability density curve of voltage of bus 27 used in Fig. 6 is the average voltage under the two models of 8760 hours in a year.

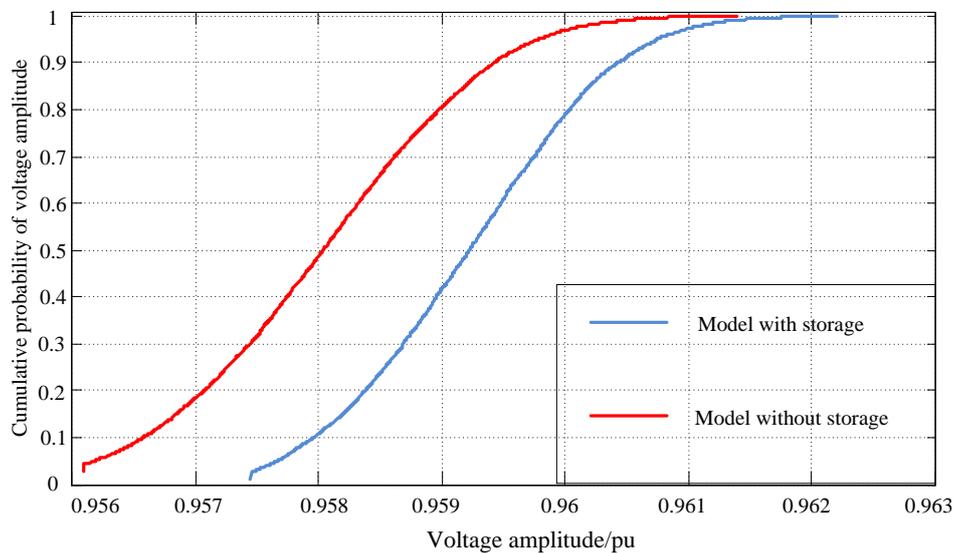

Fig. 6 Cumulative probability of the voltage amplitude at the bus 27

It can be seen from Fig. 6 that the voltage amplitude of joint planning model is higher than that of separate planning model, and the red line rises faster which shows that the voltage fluctuation becomes smaller in a year. So the joint planning model considering both DGs and energy storage in this paper has a good effect on the voltage of the network.

# 6 Conclusion

In order to reduce the planning errors caused by the uncertainties of DGs outputs, this paper establishes a joint planning model of DGs and energy storage devices by using bi-level programming for active distribution networks. Here, the upper-level model aims to seek the optimal location and capacity of DGs and energy storage, while the lower-level model optimizes the operation of energy storage devices. To solve this model, the BPSO algorithm is adopted, and the optimal planning scheme is achieved via alternating iterations between the two levels. Based on the simulation results on the PG & E 69-bus system, the following conclusions can be safely drawn:

(1) The joint planning method of DGs and storage manages to reduce the planning errors, it achieves the least annul cost compared with other methods.

(2) The daily operation optimization of the energy storage effectively alleviates the fluctuation caused by DGs, which contributes to voltage profile, peak shaving and network loss.

(3) The method proposed in this paper provides a new idea for distribution network planners in dealing with DG volatility, which makes the planning more effective and practical than traditional planning methods.

For the next step, the DGs and storage planning considering island operation will be studied, the content of economic objective function will be enriched, and a more advanced algorithm is to be adopted. In this study, the uncertainties of renewable power generations are modelled by using certain probabilistic distributions, while a more realistic modeling technique can be developed by using deep generative models based scenario generation [41]. It is also an interesting topic to leverage game theory for distribution network planning with distributed generations [42].

**Acknowledgements**

This work is partly supported by the Natural Science Foundation of Jilin Province, China under Grant No. YDZJ202101ZYTS149.